\documentstyle[12pt,epsbox]{article}

\newcommand{\gtrsim}{\mbox{\raisebox{-1.0ex}
        {$\stackrel{\textstyle ~>~}{\textstyle \sim}$}}}
\newcommand{\lesssim}{\mbox{\raisebox{-1.0ex}
        {$\stackrel{\textstyle ~<~}{\textstyle \sim}$}}}

\newcommand{\lsp}{ \left ( }
\newcommand{\rsp}{ \right ) }
\newcommand{\lmp}{ \left \{ }
\newcommand{\rmp}{ \right \} }
\newcommand{\llp}{ \left [ }
\newcommand{\rlp}{ \right ] }

\newcommand{\rabs}{ \right | }

\newcommand{\MEV}{ {\rm MeV} }
\newcommand{\GEV}{ {\rm GeV} }
\newcommand{\TEV}{ {\rm TeV} }
\newcommand{\mgra}{ m_{3/2} }
\newcommand{\epho}{ \epsilon_{\gamma} }

\newcommand{\eele}{ E_{e} }
\newcommand{\enu}{ E_{\nu} }
\newcommand{\enubg}{ \bar{E}_{\nu} }
\newcommand{\beq}{\begin{eqnarray}}
\newcommand{\eeq}{\end{eqnarray}}
\begin{document}
\baselineskip .7cm

\begin{titlepage}

\begin{flushright}
    TU-463\\
    \today
\end{flushright}
\vspace{1cm}
\begin{center}
    {\Large\bf Gravitino Decay into a Neutrino and a Sneutrino in the
    Inflationary Universe}\\
    \vspace{0.5cm}
    {\large M. Kawasaki} \\
    {\it Institute for Cosmic Ray Research, The
    University of Tokyo, Tanashi 188, Japan}\\
    {\large and}\\
    {\large T. Moroi \footnote{Fellow of the Japan Society 
    for the Promotion of Science.}}\\
    {\it Department of Physics, Tohoku University, Sendai 980, Japan}
\end{center}

\vspace{2cm}

\begin{abstract}
Gravitino produced in the inflationary universe are studied. When
the gravitino decays into a neutrino and a sneutrino, the emitted high
energy neutrinos scatter off the background neutrinos and produce
charged leptons (mainly electrons and positrons), which cause the
electro-magnetic cascades and produce many soft photons.  We obtain the
spectra of the high energy neutrinos as well as the spectrum of the high
energy photon by integrating a set of Boltzmann equations.  Requiring
these photons should not alter the abundances of the light elements (D,
$^3$He, $^4$He) in the universe, we can set the stringent upperbound on
the reheating temperature after the inflation. We find that $T_R
\lesssim (10^{10}-10^{12})$GeV for $m_{3/2}\sim (100\GEV - 1\TEV)$, which
is more stringent than the constraints in the previous works.
\end{abstract}

\end{titlepage}

\newpage

\section{Introduction}
\label{introduction}

\hspace*{\parindent}
In models based on supergravity~\cite{NPB212-413}, it has been pointed
out that there exist some light particles whose interactions are
suppressed by powers of the gravitational scale $M = M_{Pl}/\sqrt{8\pi}
\simeq 2.4\times 10^{18}\GEV$.  Such particles have nothing to do with
collider experiments, but may affect the standard scenario of big-bang
cosmology~\cite{PRL48-1303,NPB277-556,PLB131-59}. The gravitino, which
is the gauge field associated with local supersymmetry (SUSY), is one of
the weakly interacting particles in supergravity models, and we expect
that the mass of the gravitino is of order of typical SUSY-breaking
scale.

Since the gravitino has only gravitational interaction with other
particles, its lifetime is very long even if it is unstable. In standard
cosmology, the gravitino with mass $m_{3/2}\sim$ $O$(100GeV -- 10TeV) is
excluded since it decays after the big-bang nucleosynthesis (BBN) and
produces an unacceptable amount of entropy, which conflicts with the
predictions of BBN~\cite{PRL48-1303}. We may avoid this constraint by
assuming the inflationary cosmology, in which the initial abundance of
the gravitino is diluted by the exponential expansion of the
universe~\cite{PLB118-59}.  However gravitinos are reproduced by
scattering processes off the thermal radiation after the universe has
been reheated. Since the number density of the secondary gravitino is
proportional to the reheating temperature, upperbound on the reheating
temperature should be imposed not to overproduce
gravitinos~\cite{PLB127-30} -- \cite{KMI}.

In the recent paper~\cite{KMI}, we consider the unstable gravitino which
decays into photon ($\gamma$) and photino ($\tilde{\gamma}$) and derive
upperbound on the reheating temperature $T_R$ ($T_R
\lesssim 10^6\GEV$ for $m_{3/2} \sim 100\GEV$, $T_R \lesssim 10^{9}\GEV$
for $m_{3/2} \sim 1\TEV$) by requiring the produced high energy photons
should not alter the abundances of light elements synthesized in BBN.
The above constraint seems to be very stringent since such low reheating
temperature requires very small decay rate of the inflaton field. For
example, in chaotic inflation with a inflaton whose interactions are
suppressed by gravitational scale $M$, the inflaton decay rate is expected
to be $\Gamma_{inf} \sim m_{inf}^3/M_{Pl}^2$ with $m_{inf}$ being the
inflaton mass, and hence the reheating temperature is estimated
as\cite{Kolb-Terner}
\beq
T_R \sim 0.1\sqrt{\Gamma_{inf}M_{Pl}} \sim 10^{9}\GEV,
\label{tr}
\eeq
requiring that the inflaton field should produce the density
perturbations observed by COBE ($m_{inf} \sim
10^{13}\GEV$)~\cite{PRL69-3602}.

However the constraints might become much weaker when the gravitino
decays only into weakly interacting particles. In the particle content
of the minimal SUSY standard model, the only candidates are neutrino and
sneutrino.

In this letter, we assume that the gravitino decays only into a neutrino
and a sneutrino and derive constraints on the reheating temperature. In
this case, the emitted high energy neutrino may scatter off the
background neutrino and produce an electron-positron (or muon-anti-muon)
pair, which then produces many soft photons through electro-magnetic
cascade processes and destruct the light elements.  Since the
interaction between high energy neutrinos and background neutrinos is
weak, it seems that the destruction of the light elements is not
efficient.  In fact, Gratsias, Scherrer and Spergel~\cite{PLB262-198}
showed that the constraint is not so stringent for the case that the
gravitino decays into a neutrino and a sneutrino.  However the previous
analysis seems to be incomplete in a couple of points.  First Gratsias
et.al.~\cite{PLB262-198} totally neglect the secondary high energy
neutrinos which are produced by neutrino-neutrino scattering. The effect
of the secondary neutrino may be important for heavy gravitino ($m_{3/2}
\gtrsim 1$TeV).  Second, they only study the case where the destruction
of $^4$He results in the overproduction of $(^3{\rm He} + {\rm D})$.
However, for the heavy gravitino which decays in early stage of the BBN,
the destruction of D is more important since the electro-magnetic
cascade process is so efficient that the energy of soft photons becomes
less than the threshold of $^4$He.  Furthermore, as pointed out in
ref.\cite{KMI}, the previous estimation of the gravitino production in
the reheating epoch is underestimated. Those effects which are not taken
into account in ref.\cite{PLB262-198} may lead to more stringent
constraint on the reheating temperature.  Therefore, in this letter, we
reexamine the decay of the gravitino into a neutrino and a sneutrino
($\psi_{\mu} \rightarrow \nu + \tilde{\nu}$) with taking all relevant
effect into account, and as a result, we have obtained more stringent
upperbound on the reheating temperature.

\section{Neutrino-Neutrino Interaction}
\label{sec:neutrino}

\hspace*{\parindent}
The high energy neutrinos ($\nu$) produced in the gravitino decay scatter off
the thermal neutrino ($\nu_b$) in the background by the following processes;
\begin{eqnarray}
    \nu_{i} + \nu_{i,b} &\rightarrow & \nu_i + \nu_i ,\\
    \nu_{i} + \bar{\nu}_{i,b} &\rightarrow &\nu_i + \bar{\nu}_i,\\
    \nu_{i} + \bar{\nu}_{i,b} &\rightarrow &\nu_j + \bar{\nu}_j ,\\
    \nu_{i} + \nu_{j,b} &\rightarrow &\nu_i + \nu_j,\\
    \nu_{i} + \bar{\nu}_{j,b} & \rightarrow &\nu_i + \bar{\nu}_j,\\
    \nu_{i} + \bar{\nu}_{i,b} &\rightarrow & e^{-}  + e^{+},\\
    \nu_{i} + \bar{\nu}_{i,b} & \rightarrow & \mu^{-}  + \mu^{+},
\end{eqnarray}
where index $i$ and $j$ represent $e$, $\mu$ and $\tau$ with $i\neq j$.
All the amplitude square $|{\cal M}|^2$ in these reactions take the form
given by
\beq
    |{\cal M}|^2 = 32G_F^2 
    \lmp a (pp')^2 + b(pq)^2 + c(pq')^2 + d m^2 (pp') \rmp ,
    \label{amplitude}
\eeq
where $G_F \simeq 1.17\times 10^{-5}\GEV^{-2}$ is the Fermi constant,
the coefficients $a$ -- $d$ depend on the individual reaction, $p$ and
$p'$ are the initial momenta of high energy neutrino and background
neutrino, $q$ and $q'$ are the final momenta, and $m$ represents the
mass of the fermion the in final state. Then one can obtain the
Boltzmann equations describing the evolution of the spectra for the high
energy neutrinos;
\begin{eqnarray}
    \frac{\partial f_{\nu_i}(\enu^\prime)}{\partial t} &=& 
    \frac{4G_F^2}{3\pi}
    \int_{\enu^\prime}^{\infty} \frac{d \enu}{\enu^2}
    \sum_j \lmp a_{in,ij}\enu^2 
    + b_{in,ij}(\enu - \enu^\prime)^2 
    + c_{in,ij} \enu^{\prime 2} \rmp f_{\nu_j}(\enu)
    \nonumber \\ 
    & & \times \int_0^{\infty} d \enubg \enubg
    \bar{f}_{\nu}(\enubg)
    \nonumber \\ 
    & & -\frac{4G_F^2}{3\pi} \enu^\prime f_{\nu_i}(\enu^\prime) 
    \lsp a_{out} + \frac{1}{3}b_{out} + \frac{1}{3}c_{out} \rsp
    \int_0^{\infty}d\bar{\epsilon}_{\nu} \bar{\epsilon}_{\nu} 
    \bar{f}_{\nu}(\bar{\epsilon}_{\nu}) 
    \nonumber\\ 
    & & + \left(
    \frac{\partial f_{\nu_i}(\enu^\prime)}
    {\partial t}\right)_{\nu_i + \bar{\nu}_i
    \rightarrow e^{-}+e^{+}} 
    + \left(\frac{\partial f_{\nu_i}(\enu^\prime)}
    {\partial t}\right)_{\nu_i + \bar{\nu}_i \rightarrow
    \mu^{-}+\mu^{+}}
    \nonumber\\ 
    & & + 
    \frac{1}{6\tau_{3/2}}n_{3/2}
    \delta(\enu^\prime -m_{3/2}/2)
    \nonumber \\
    & & + \enu^\prime H \frac{\partial f_{\nu_i}(\enu^\prime)}
    {\partial \enu^\prime}
    - 2 H f_{\nu_i}(\enu^\prime)\enu^\prime,
\label{df_nu/dt}
\end{eqnarray}
where $H$ is the expansion rate of the universe, coefficients $a$ -- $c$
are given by
\begin{eqnarray}
    &&
    a_{out} = 4,~~~b_{out} = 0,~~~c_{out} = 13,
    \\ &&
    a_{in,ii}= 6,~~~b_{in,ii}= 9,~~~c_{in,ii}= 11,
    \\ &&
    a_{in,ij}= 1,~~~b_{in,ij}= 1,~~~c_{in,ij}= 2,~~~(i\neq j),
\end{eqnarray}
and the spectrum of the background neutrino $\bar{f}_{\nu}$ can be
written as
\begin{eqnarray}
    \bar{f}_{\nu} (\enubg) =\frac{\enubg^2}{2\pi^2}
    \frac{1}{e^{\enubg / T_{\nu}}+1},
\end{eqnarray}
with $T_{\nu}$ being the neutrino temperature. The last two terms of
right-hand side of eq.(\ref{df_nu/dt}) represent the effect of the cosmic
expansion. The formula for charged lepton pair creation is obtained as
\beq
\lsp \frac{f_{\nu_i}(\enu)}{\partial t}
\rsp_{\nu_i + \bar{\nu}_i \rightarrow l^{-}+l^{+}} &=&
-\frac{G_F^2}{16\pi} \frac{1}{\enu^2} f_{\nu_i}(\enu)
\int_0^\infty d \enubg \bar{f}_\nu (\enubg)
\nonumber  \\ &&
\times \lmp \lsp a + \frac{1}{3} b + \frac{1}{3} c \rsp I_2 +
\lsp 2d - \frac{1}{3} b -  \frac{1}{3} c \rsp m^2  I_1 \rmp,
\eeq
with
\beq
I_2 &=& \frac{4}{3} \lmp 4 - \frac{4m^2}{\enu \enubg} \rmp^{1/2}
\enu \enubg \lsp 8 \enu^2 \enubg^2 - 2 m^2 \enu \enubg - 3m^4 \rsp
\nonumber \\ &&
- 4m^6 \ln \llp
\frac{2 \lmp 4 - \lsp 4m^2/\enu \enubg \rsp \rmp^{1/2} \enu \enubg
+ 4 \enu \enubg - 2m^2}{2m^2} \rlp,
\\
I_1 &=& 2 \lmp 4 - \frac{4m^2}{\enu \enubg} \rmp^{1/2}
\enu \enubg \lsp 2 \enu \enubg - m^2 \rsp
\nonumber \\ &&
- 2m^4 \ln \llp
\frac{2 \lmp 4 - \lsp 4m^2/\enu \enubg \rsp \rmp^{1/2} \enu \enubg
+ 4 \enu \enubg - 2m^2}{2m^2} \rlp.
\eeq
Coefficients for the charged lepton production processes are given by
\beq
a_{\nu_i + \bar{\nu}_i \rightarrow l_i^{-}+l_i^{+}} &=& 0,
\\
b_{\nu_i + \bar{\nu}_i \rightarrow l_i^{-}+l_i^{+}} &=& 16\sin^4\theta_W,
\\
c_{\nu_i + \bar{\nu}_i \rightarrow l_i^{-}+l_i^{+}} &=& 16\sin^4\theta_W,
\\
d_{\nu_i + \bar{\nu}_i \rightarrow l_i^{-}+l_i^{+}} &=& 16\sin^4\theta_W,
\\
a_{\nu_i + \bar{\nu}_i \rightarrow l_j^{-}+l_j^{+}} &=& 0,
\\
b_{\nu_i + \bar{\nu}_i \rightarrow l_j^{-}+l_j^{+}} &=& 16\sin^4\theta_W,
\\
c_{\nu_i + \bar{\nu}_i \rightarrow l_j^{-}+l_j^{+}} &=& 
\lsp 16\sin^2\theta_W - 2 \rsp^2 , 
\\
d_{\nu_i + \bar{\nu}_i \rightarrow l_j^{-}+l_j^{+}} &=& 
16\sin^4\theta_W - 8 \sin^2\theta_W,
\eeq
with $\theta_W$ being the Weinberg angle and $i\neq j$. The gravitino
decay rate $\Gamma_{3/2}$ and the number density $n_{3/2}$ of the
gravitino are given by
\begin{equation}
    \Gamma_{3/2}(\psi_{\mu} \rightarrow \nu + \tilde{\nu})
    = \sum_{i} \frac{m_{3/2}^3}{192 \pi M}
    \lmp 1 - \lsp \frac{m_{\tilde{\nu}_i}}{m_{3/2}} \rsp^2 \rmp^4,
    \label{lifetime}
\end{equation}
and\footnote
{In the recent article, Fischler~\cite{PLB332-277} have proposed a new
mechanism to produce gravitinos in thermal bath. If one adopts his
mechanism, the constraints obtained below must become more stringent. 
However, it is not clear to us if his new mechanism is relevant.}
\begin{equation}
    n_{3/2} \simeq  1.07 \times 10^{-11} n_{\gamma}
    \left(\frac{T_R}{10^{10}{\rm GeV}} \right)
    \left[ 1 -0.0232\log\left(\frac{T_R}{10^{10}{\rm GeV}}\right)\right]
    e^{-\Gamma_{3/2}t}
    \label{abundance}
\end{equation}
where $m_{\tilde{\nu}_i}$ is the mass of the sneutrino of $i$-th
generation and $n_{\gamma}$ is the photon number density. In the
following analysis, we assume all the sneutrino masses are degenerate
for simplicity.

In Fig.\ref{fig:nu-spec} the spectra of three types of neutrinos at
$T=1$eV are shown for $m_{3/2} = 100$GeV, 1TeV and 10TeV.  As one can
see, the differences among three types of neutrinos are very small since
the differences only come from the charged lepton pair creation
processes which are sub-dominant compared with the neutrino-neutrino
scattering processes.  For the heavy gravitino ($m_{3/2} \gtrsim 1$TeV)
the decay occurs early in the universe when the neutrino temperature is
so high that the $\nu-\nu_b$ scatterings is effective. (Notice that the
cross section is roughly given by $\langle\sigma v\rangle \sim G_F^2
E_{\nu}T_{\nu}$.)  Therefore high energy neutrinos emitted from the
gravitinos scatter off the background neutrinos frequently and produce
many secondary neutrinos. On the other hand, the emitted neutrinos
seldom interact with the background neutrinos if the gravitino mass is
small ($m_{3/2} \lesssim 100$GeV) and the effect of the neutrino-neutrino
scattering is negligible.

\section{High Energy Photon Spectrum}
\label{sec:Photon spectrum}

\hspace*{\parindent}
The primary and secondary high energy neutrinos scatter off the
background neutrinos and produce charged lepton pairs. Then the charged
leptons induce the electro-magnetic cascade processes.  In the same
procedure as ref.\cite{KMI} we obtain the high energy photon and
electron spectra, taking the following radiative processes into account.
(I) High energy photon with energy $\epho$ can become electron-positron
pair by scattering off background photon if the energy of the background
photon is larger than $m_{e}^{2}/\epho$ where $m_{e}$ is the electron
mass. We call this process double photon pair creation. For sufficiently
high energy photons, this is the dominant process since the cross
section or the number density of target is much larger than other
processes.  Numerical calculation shows that this process determines the
shape of the spectrum of high energy photon for $\epho\gtrsim
m_{e}^{2}/22T$.  (II) Below the effective threshold of double photon
pair creation, high energy photons lose their energy by photon-photon
scattering. But in the limit of $\epho \rightarrow 0$, the total cross
section for the photon-photon scattering is proportional to $\epho^{3}$
and this process loses its significance.  Hence finally, photons are
thermalized by (III) pair creation in nuclei, or (IV) Compton scattering
off background electron. And (V) emitted high energy electrons and
positrons lose their energy by the inverse Compton scattering off
background photon.

The Boltzmann equations for the photon and electron distribution
function $f_{\gamma}$ and $f_{e}$ are given by
\begin{eqnarray}
\frac{\partial f_{\gamma}(\epho)}{\partial t}
&=&
\left. \frac{\partial f_{\gamma}(\epho)}{\partial t} \right |_{\rm DP}
+ \left. \frac{\partial f_{\gamma}(\epho)}{\partial t} \right |_{\rm PP}
+ \left. \frac{\partial f_{\gamma}(\epho)}{\partial t} \right |_{\rm PC}
\nonumber \\ &&
+ \left. \frac{\partial f_{\gamma}(\epho)}{\partial t} \right |_{\rm CS}
+ \left. \frac{\partial f_{\gamma}(\epho)}{\partial t} \right |_{\rm IC},
\label{bol-fp} \\
\frac{\partial f_{e}(\eele)}{\partial t}
&=&
\left. \frac{\partial f_{e}(\eele)}{\partial t} \right |_{\rm DP}
+ \left. \frac{\partial f_{e}(\eele)}{\partial t} \right |_{\rm PC}
+ \left. \frac{\partial f_{e}(\eele)}{\partial t} \right |_{\rm CS}
\nonumber\\ &&
+ \left. \frac{\partial f_{e}(\eele)}{\partial t} \right |_{\rm IC}
+ \left. \frac{\partial f_{e}(\eele)}{\partial t} \right |_{\rm NEU},
\label{bol-fe}
\end{eqnarray}
where DP (PP, PC, CS, IC, and NEU) represents double photon pair
creation (photon-photon scattering, pair creation in nuclei, Compton
scattering, inverse Compton scattering, and the contribution from the
$\nu - \nu$ scatterings).  In solving eqs.(\ref{bol-fp}) and
(\ref{bol-fe}), we assume
\beq
\left. \frac{f_{e}(E)}{\partial t} \rabs_{\rm NEU}=
- \sum_i \lsp 
\left. \frac{f_{\nu_i}(E)}{\partial t} 
\rabs_{\nu_i + \bar{\nu}_i \rightarrow e^{-}+e^{+}} + 
\left. \frac{f_{\nu_i}(E)}{\partial t} 
\rabs_{\nu_i + \bar{\nu}_i \rightarrow \mu^{-}+\mu^{+}} \rsp,
\eeq
which respects the energy conservation although the actual energy
distribution may be different. This assumption is adequate for our
purpose since the photon (and electron) spectrum is determined almost
only by the total amount of the energy injection~\cite{KMI}.
Furthermore we treat muons (and anti-muons) as electrons with the same
energy and neglect the contribution from tau leptons whose creation rate
is much smaller than the other charged leptons.  Full details for other
terms are shown in ref.\cite{KMI}.

\section{BBN and Photo-Dissociation of Light Elements}

\hspace*{\parindent}
The presence of gravitino might destroy the success of BBN.  The
gravitino may have three effects on BBN. First the energy density of the
gravitino at $T\simeq 1$MeV speeds up the cosmic expansion and leads to
increase the $n/p$ ratio at the neutron decoupling time and hence $^4$He
abundance also increases.  Second, the entropy production due to the
decay of the gravitino reduces the baryon-to-photon ratio and results
in too baryon-poor universe.  Third, the high energy photons induced by
the decay of the gravitino destroy the light elements (D, $^3$He,
$^4$He).  Among three effects, photo-dissociation of the light elements
by high energy photon is the most important for the relatively light
gravitino ($m_{3/2}\lesssim$ 1TeV). The entropy production is not
important for the present case since gravitinos decay into invisible
particles (i.e.  neutrinos).
 
When the energy of the high energy photon is relatively low, i.e.
$2\MEV \lesssim \epho \lesssim 20\MEV$, the D, T and $^3$He are
destroyed and their abundances decrease. On the other hand, if the
photons have high energy enough to destroy $^4$He, it seems that such
high energy photons only decrease the abundance of all light elements.
However since D, T and $^3$He are produced by the photo-dissociation of
$^4$He whose abundance is much higher than the other elements, their
abundances can increase or decrease depending on the number density of
the high energy photon. When the number density of the high energy
photons with energy greater than $\sim$ 20MeV is extremely high, all
light elements are destroyed. But as the photon density becomes lower,
there is some range of the high energy photon density at which the
overproduction of D, T and $^3$He becomes significant.  And if the
density is sufficiently low, the high energy photon does not affect the
BBN at all.

From various observations, the primordial abundances of light elements
are estimated~\cite{Walker} as 
\begin{eqnarray}
    &&  0.22 < Y_p \equiv 
    \lsp \frac{\rho_{^{4}{\rm He}}}{\rho_{B}}\rsp_p < 0.24 ,
    \label{obs-he4}\\[0.5em]
    && \lsp \frac{n_{\rm D}}{n_{\rm H}} \rsp_p > 1.8\times 10^{-5},
    \label{obs-h2}\\[0.5em]
    && \lsp\frac{n_{\rm D}+n_{^3{\rm He}}}{n_{\rm H}} \rsp_p
    <  1.0 \times 10^{-4},
    \label{obs-h23}
\end{eqnarray}
where ${\rho_{^{4}{\rm He}}}$ and ${\rho_{B}}$ are the mass densities of
$^4$He and baryon. The abundances of light elements modified by the
gravitino decay must satisfy the observational constraints above. In
order to make precise predictions for the abundances of light elements,
we have modified Kawano's computer code~\cite{Kawano} to include the
photo-dissociation processes.

In the present calculation there are at least three free parameters,
i.e. the mass of the gravitino $m_{3/2}$, reheating temperature $T_R$ and
the baryon-to-photon ratio $\eta_B$.  However as shown in the previous
paper~\cite{KMI}, the baryon-to-photon ratio $\eta_B$ is not important
parameter because the allowed value for $\eta_{B}$ is almost the same
as that in the standard case (i.e. without gravitino). Therefore we
fix $\eta_B = 3\times 10^{-10}$ in this letter.

\section{Results and Discussions}
\label{sec:results}

\hspace*{\parindent}
The allowed regions that satisfy the observational constraints
(\ref{obs-he4})--(\ref{obs-h23}) is shown in the $m_{3/2}-T_R$ plane in
Fig.\ref{fig:const}. In Fig.\ref{fig:const} one can see that for the
gravitino with mass between $\sim$ 100GeV and $\sim$ 1TeV, the
overproduction of ${\rm D}$ and $^3{\rm He}$ gives the most stringent
constraint, while the upperbound on the reheating temperature is
determined from the destruction of D for $m_{3/2} \simeq (1-3)$TeV.
Notice that D destruction is not considered in the previous
work~\cite{PLB262-198}.  Furthermore the constrain from $({\rm D} +
^3{\rm He})$ overproduction is more stringent. The reasons why we obtain
the more stringent constrain are i) the gravitino abundance is (4--5)
times higher than the one obtained in ref.\cite{PLB145-181} that the
previous authors used, ii) the secondary neutrinos are taken into
account in our calculation, and iii) the gravitino lifetime for
$\psi_{\mu} \rightarrow
\nu + \tilde{\nu}$ is longer by a factor 2 than the lifetime for $\psi_{\mu}
\rightarrow \gamma + \tilde{\gamma}$. (In ref.\cite{PLB262-198} it is 
presumed that $\Gamma_{3/2}(\psi_{\mu} \rightarrow \nu +\tilde{\nu}) =
\Gamma_{3/2}(\psi_{\mu} \rightarrow \gamma + \tilde{\gamma})$.)

In Fig.\ref{fig:const}, the overproduction of $^4$He due to the mass
density of the gravitino at $T\sim 1$MeV is also seen. The constraint
from this effect is less stringent than those from the
photo-dissociation of D and the present sneutrino mass density
(discussed below).

If we assume that the sneutrino is the lightest SUSY particle, the
sneutrinos produced by the gravitino decay are stable and their
present mass density is given by
\begin{equation}
    \Omega_{\tilde{\nu}}h^2 = 0.037 
    \left(\frac{m_{\tilde{\nu}}}{\rm 100GeV}\right)
        \left(\frac{T_R}{10^{10}\GEV}\right),
    \label{density}
\end{equation}
where $h$ is the Hubble constant in units of 100km/sec/Mpc.  Therefore
the present limit on sneutrino mass 41GeV~\cite{PD} sets the upperbound
on the reheating temperature;
\begin{eqnarray}
    T_{R} \leq 6.6\times 10^{11}  h^{2}~\GEV.
    \label{ub-omega}
\end{eqnarray}
This constraint is also shown in Fig.\ref{fig:const}.\footnote
{In calculating eq.(\ref{density}), we have ignored the effect of the
pair annihilation of sneutrinos ($\tilde{\nu}+\tilde{\nu}^* \rightarrow
f + \bar{f}$). For the pair annihilation process, the cross section is
roughly estimated as $\sigma v\lesssim\alpha_W/m_{\tilde{\nu}}^2$ (where
$\alpha_W$ is the coupling factor and $v$ is the relative velocity), and
the condition for sufficiently large pair annihilation rate
($n_{\tilde{\nu}}\sigma v \gtrsim H$, with $n_{\tilde{\nu}}$ being the
sneutrino density) reduces to $n_{\tilde{\nu}}/n_{\gamma}\gtrsim
10^{-7}$ (with $Y_{\tilde{\nu}}$ being the yield variable for the
sneutrino), which is less stringent than the constraint
(\ref{ub-omega}). However, if the sneutrino mass is small enough to hit
the pole of the $Z$-boson propagator, the constraint (\ref{ub-omega})
becomes weaker, and constraint from the $^4$He overproduction may become
significant for large gravitino mass ($\mgra\gtrsim 3\TEV$).}

In summary, we have investigated the decay of the gravitino into a
neutrino and a sneutrino, in particular, the effect on BBN by high
energy photons induced by the decay. We have found the constraint on
reheating temperature and mass of the gravitino as
\begin{equation}
    T_{R} \lesssim 10^{10-12}\GEV  ~~~~~~ {\rm for}
    ~~~~ m_{3/2} =100\GEV - 10\TEV,
\end{equation}
which is more stringent than the one obtained in ref.\cite{PLB262-198},
but compatible with the rough estimation of the reheating temperature
(\ref{tr}) for a chaotic inflation. 

\subsection*{Acknowledgement}

We would like to thank T. Yanagida for useful discussions, and to K.
Maruyama for informing us of the experimental data for
photo-dissociation processes. One of the author (T.M.) thanks Institute
for Cosmic Ray Research where part of this work has done.  This work is
supported in part by the Japan Society for the Promotion of Science.

\newpage
%

%
%
\newcommand{\Journal}[4]{{\sl #1} {\bf #2} {(#3)} {#4}}
\newcommand{\APJ}{\sl Ap. J.}
\newcommand{\CJP}{\sl Can. J. Phys.}
\newcommand{\NC}{\sl Nuovo Cimento}
\newcommand{\NP}{\sl Nucl. Phys.}
\newcommand{\PL}{\sl Phys. Lett.}
\newcommand{\PR}{\sl Phys. Rev.}
\newcommand{\PRL}{\sl Phys. Rev. Lett.}
\newcommand{\PTP}{\sl Prog. Theor. Phys.}
\newcommand{\SJNP}{\sl Sov. J. Nucl. Phys.}
\newcommand{\ZP}{\sl Z. Phys.}

\newpage

\begin{center}
    {\large\bf Figure Captions}
\end{center}

\begin{figure}[htbp]
\caption{Spectra of high energy electron-neutrino (solid curve),
muon-neutrino (dashed curve) and tau-neutrino (dotted curve) at $T=1$eV
for $m_{3/2} = 100\GEV$, $1\TEV$ and $10\TEV$. We have taken $T_R
=10^{10}$GeV.} \label{fig:nu-spec}
\end{figure}

\begin{figure}[htbp]
\caption{Constraints on the reheating temperature from BBN, the
present mass density of the sneutrino. The region above the curves is
excluded. The lower mass of the gravitino from $m_{3/2} > m_{\tilde{\nu}} >
41$GeV is also shown in long-short dashed line. } \label{fig:const}
\end{figure}

%
%
\begin{figure}[t]
\begin{center}

\epsfile{file=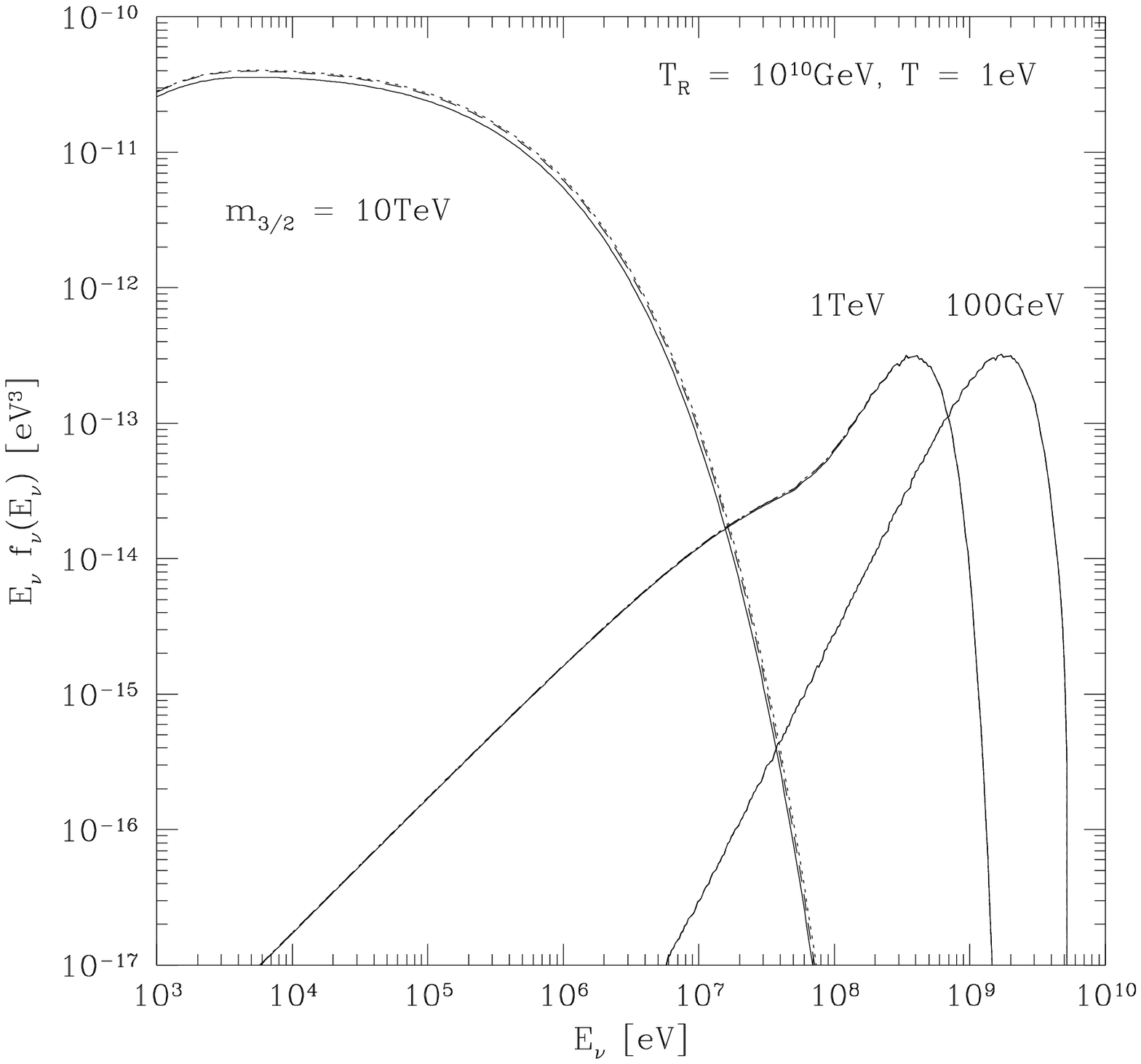,width=16cm}

\vspace{2cm}

{\bf Fig. \protect\ref{fig:nu-spec}}

\end{center}
\end{figure}

%
%
\begin{figure}[t]
\begin{center}

\epsfile{file=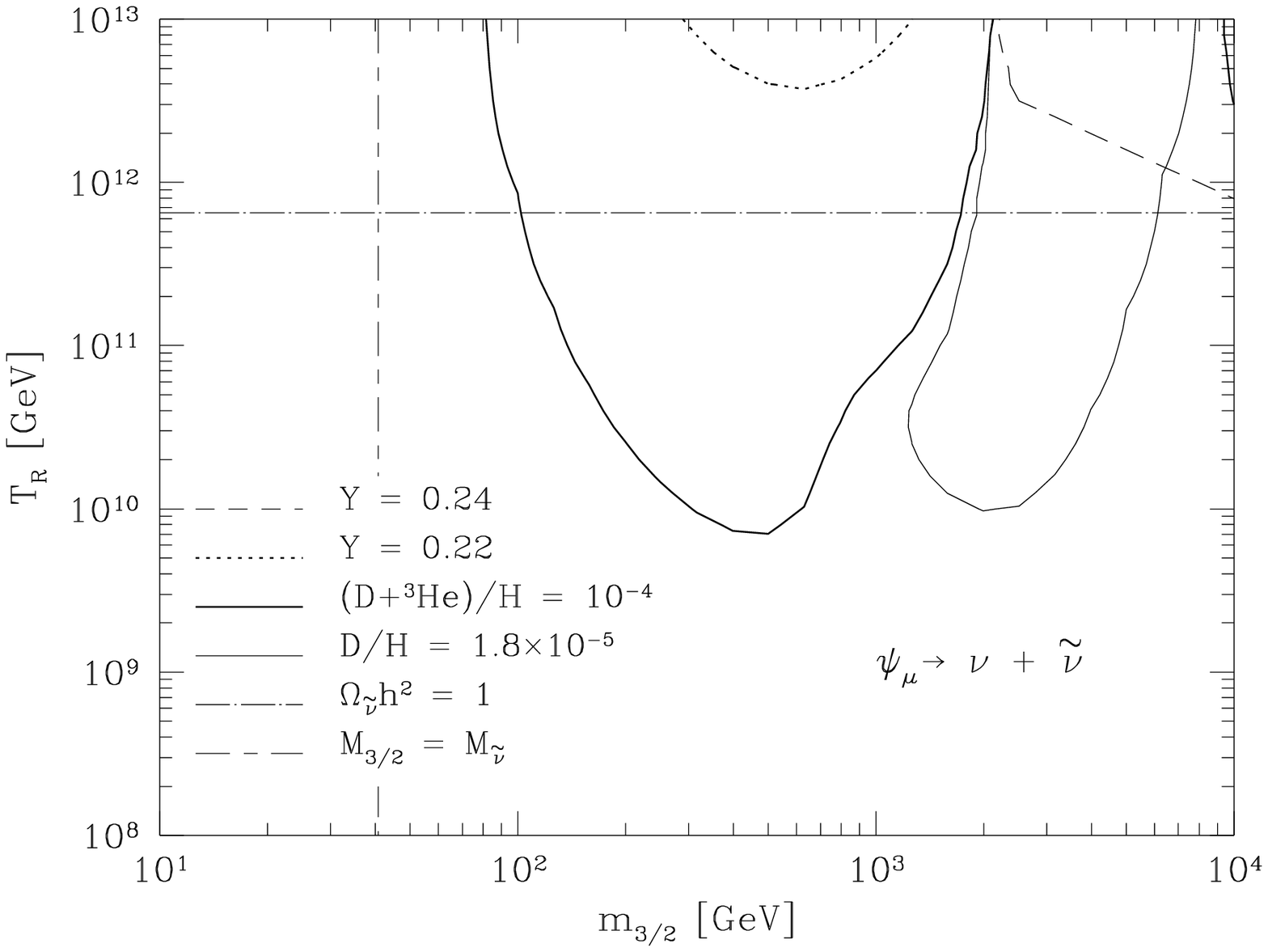,width=16cm}

\vspace{2cm}

{\bf Fig. \protect\ref{fig:const}}

\end{center}
\end{figure}

\end{document}